\documentclass[twocolumn, times]{aastex631}
\usepackage{mathptmx}
\usepackage{fancyvrb}
\usepackage{epstopdf}

\usepackage{amsmath}
\usepackage{mathtools}
\usepackage{amssymb}

\usepackage{graphicx}
\usepackage{natbib}
\usepackage{footnote}
\usepackage{bbold}
\usepackage{paralist}
\usepackage{bbm}
\usepackage{import}

\newcommand{\new}[1]{ { #1} }
\DeclarePairedDelimiter{\nint}\lfloor\rceil

\graphicspath{{./}{plots/}}
\begin{document}

\title{Recovering Pulsar Periodicity from Time-of-Arrival Data by Finding the Shortest Vector in a Lattice}


\author{Dotan Gazith}
\affiliation{Department of Particle Physics and Astrophysics, Weizmann Institute of Science, 76100 Rehovot, Israel}
\email{dotan.gazith@weizmann.ac.il}

\author[0000-0002-8912-0732]{Aaron~B.~Pearlman}
\altaffiliation{Banting Fellow, McGill Space Institute~(MSI) Fellow, \\ and FRQNT~Postdoctoral Fellow.}
\affiliation{Department of Physics, McGill University, 3600 rue University, Montr\'eal, QC H3A 2T8, Canada}
\affiliation{Trottier Space Institute, McGill University, 3550 rue University, Montr\'eal, QC H3A 2A7, Canada}
\affiliation{Division of Physics, Mathematics, and Astronomy, California Institute of Technology, Pasadena, CA 91125, USA}
\email{aaron.b.pearlman@physics.mcgill.ca}

\author{Barak Zackay}
\affiliation{Department of Particle Physics and Astrophysics, Weizmann Institute of Science, 76100 Rehovot, Israel}
\email{barak.zackay@weizmann.ac.il}

\begin{abstract}
The strict periodicity of pulsars is one of the primary ways through which their nature and environment can be studied, and it has also enabled precision tests of general relativity and studies of nanohertz gravitational waves using pulsar timing arrays~(PTAs). Identifying such a periodicity from a discrete set of arrival times is a difficult algorithmic problem, particularly when the pulsar is in a binary system. This challenge is especially acute in $\gamma$-ray pulsar astronomy, as there are hundreds of unassociated Fermi-LAT sources that may be produced by $\gamma$-ray emission from unknown pulsars. Recovering their timing solutions will help reveal their properties and may allow them to be added to PTAs.
The same issue arises when attempting to recover a strict periodicity for repeating fast radio bursts~(FRBs). Such a detection would be a major breakthrough, providing us with the FRB source's age, magnetic field, and binary orbit. The problem of recovering a timing solution from sparse time-of-arrival (TOA) data is currently unsolvable for pulsars in unknown binary systems and incredibly hard even for isolated pulsars. In this paper, we frame the timing recovery problem as the problem of finding a short vector in a lattice and obtain the solution using off-the-shelf lattice reduction and sieving techniques. As a proof of concept, we solve PSR~J0318+0253, a millisecond $\gamma$-ray pulsar discovered by FAST in a $\gamma$-ray directed search, in a few CPU-minutes.
We discuss the assumptions of the standard lattice techniques and quantify their performance and limitations.

\end{abstract}

\section{Introduction}
\label{sec:Introduction}

The pulsar search problem, recovering the timing parameters of a previously unknown pulsar, is central to pulsar astronomy. Timing pulsars allow us to learn about their age, magnetic field, and formation scenario \citep{HandbookPulsar}. After obtaining an initial timing solution, precision pulsar timing allows the use of pulsars as tools to study GR \citep{PulsarTimingGR}, galactic and globular cluster dynamics \citep{PulsarTimingGlobularCluster}, and the gravitational wave background \citep{IPTA2022}. 
When the observations are grouped together and the effective rotational period can be measured on short time scales (for example, radio observations of pulsars), solving the timing problem to phase connect all observations is not computationally demanding (though sometimes nontrivial; \citealt{AlgorithmicPulsarTiming}).

Periodicity searches become extremely challenging when the data consist of a small set of sparsely spaced times of arrival (TOAs) or phase measurements, as the number of distinct possible timing parameter values rapidly increases with the observation's duration.
Although solving this problem has been important for several decades, it has not yet been solved. However, fruitful efforts have been made using semi-coherent techniques, which can, to some extent, also tackle the fully coherent search problem \citep{AtwoodTimeDifferencing, PletchSemiCoherent, NiederSemiCoherentCircular, GPUAcceleratedEinsteinAtHome}.
As a result, more than a thousand Fermi-LAT unassociated point sources might be pulsars \citep{4FGL-DR4}, where machine learning-based studies have predicted 100--700 of them may be pulsars \citep{MLclass4, MLclass1, MLclass3, MLclass2}.

With existing techniques, obtaining a timing solution for a single millisecond pulsar (MSP) for a Fermi-LAT unassociated source is an extremely demanding computational task despite using substantial computational resources. Semi-coherent algorithms are often used, which require substantial computational resources and compromise on sensitivity \citep{AtwoodTimeDifferencing, PletchSemiCoherent, NiederSemiCoherentCircular, GPUAcceleratedEinsteinAtHome}. This scheme is highly suboptimal when applied to solving for MSPs in binary systems (most MSPs are formed through a ``recycling'' process, where the pulsar's high rotational frequency originates from matter being accreted from a companion in a binary system; \citealt{MSPsBinaryLorimer}).

Extensive distributed volunteer community efforts (e.g., \textit{Einstein@Home};~\citealt{Allen+2013}) have been made to bypass this algorithmic difficulty, and many unassociated Fermi-LAT point sources have been blindly followed up using state-of-the-art radio facilities (e.g., \citealt{MeerKATFGLFollowup, FermiRadioFollowupBruzewski, FrailFermiRadioFollowup}). Other techniques, such as cross-matching Fermi-LAT sources with optical sources exhibiting periodic modulation, have enabled the recovery of several of the timing parameters (e.g., precise position, proper motion, and orbital period), reducing the computational load by more than 8 orders of magnitude and making the recovery effort feasible with current algorithms~\citep{SearchOpticalFermiLAT}. 

Another outstanding example of the need for an algorithmic solution to the pulsar search problem is the effort to search for a strict periodicity in the arrival times of repeating fast radio bursts~(FRBs). FRBs are a class of extragalactic astrophysical sources, characterized by extremely luminous radio bursts~\citep{Petroff+2019, Cordes+2019, Bailes2022}, with durations ranging from nanoseconds to milliseconds~\citep{Majid+2021, Nimmo+2022, Snelders+2023}. These radio bursts have a wide range of applications, including cosmological studies (e.g., see~\citealt{FRBCosmology, Macquart+2020}), understanding the FRB emission engine~(e.g., see~\citealt{CHIME+2020b, Pearlman+2023}), and distinguishing between different source types~(e.g., see~\citealt{Kirsten+2022, Bhardwaj+2023a}). Some FRBs have been observed to emit multiple bursts and are referred to as repeating FRBs (e.g., see~\citealt{CHIME+2019b, Fonseca+2020, CHIME+2023a}). It is still unclear if all FRBs repeat and what are the burst emission statistics~\citep{FRBReview}.

An important hint as to why this search is expected to be computationally hard is the discovery of many days periodicity in the activity of some repeating FRBs, hinting at a binary origin~\citep{CHIME+2020b}. The discovery of a sub-second periodicity in the so-far non-repeating FRB~20191221A suggests that some FRB sources may be powered by rotating neutron stars with periodic radio emission~\citep{CHIME+2022a}. However, a timing solution from a repeating FRB still remains elusive, despite substantial search efforts and the detection of hundreds of bursts from several sources~\citep{FRB121102FAST, Xu+2022, Niu+2022, FRB_PER_SEARCH}.

A leading candidate for the FRB engine is a highly magnetized neutron star~\citep{FRBTheoryCat}, based on the short duration of the observed radio emission and the similar phenomenological characteristics shared with pulsars~\citep{Pearlman+2018a}. Nearly all phenomena (including giant pulses) related to neutron stars have temporal properties that reveal the neutron star rotation \citep{GiantBurstsCrab}. If repeating FRBs also share this property, it is, in principle, possible to to obtain a timing model using the arrival times of the radio bursts, where the arrival times of the bursts would cluster in rotational phase, similar to giant pulses from the Crab pulsar \citep{GiantBurstsCrab}. 

Pulsar timing models can be incredibly precise and sensitive to many significant digits in the rotation frequency, frequency derivatives, sky position, and Keplerian (and post-Keplerian) orbital parameters. Measuring all the above parameters for a repeating~FRB would allow us to study their astrophysical formation scenario through measurements of their age, surface magnetic field, orbital period, eccentricity, and binary mass function. Since FRBs are very extreme systems, orders of magnitude brighter than other known Galactic pulsars 
\citep{FRBReview}, their formation scenario may reveal rare, yet important, phenomena related to the formation of compact objects and perhaps even their influence on their surroundings.

\subsection{The Pulsar Search Problem and Existing Solutions}
The observed arrival times of the bursts, $t_{\rm obs}$, can be modeled as:
\begin{align}
    t_{\rm obs} = t_{\rm em} + \Delta t_{\rm orb} + \Delta t_{\rm prop},
\end{align}
where $t_{\rm em}$ is the emission time of the burst, $\Delta t_{\rm orb}$ is the delay due to the orbital motion of the source, and $\Delta t_{\rm prop}$ is the delay due to the propagation in the solar system (geometric and relativistic corrections).
A perfectly periodic source satisfies:
\begin{align}
    t_{\rm em} \bmod P_{\rm rot} = \phi_t.
\end{align}
Equivalently, we can write 
\begin{align}
f_{\rm rot} t_{\rm em} = K + \phi,
\end{align} where $f_{\rm rot}$ is the source's rotational frequency and $K$ is an \emph{integer}.
A source whose rotation rate is changing linearly with time satisfies:
\begin{align}
    f_{\rm rot} t_{\rm em} + \dot{f}_{\rm rot} \frac{t^2_{\rm em}}{2} = K + \phi.
\end{align}
To characterize the computational hardness of finding the timing solution, we can estimate the number of ``independent'' timing models that we would need to enumerate in a brute-force search for the timing model by:
\begin{align}
    \Lambda \equiv N_{\rm rot} N_{\rm geom} N_{\rm orb},
\end{align}
where $N_{\rm rot} \equiv N_f \times N_{\dot{f}}$, and $N_f \equiv \frac{f}{\delta_f}$ and $N_{\dot{f}}\equiv \frac{\dot{f}}{\delta_{\dot{f}}}$, where $\delta_f$ and $\delta_{\dot{f}}$ are the typical measurement errors in the timing model.  
Similarly (but perhaps with more complications),
$N_{\rm geom}$ is the characteristic number of options for ``independent'' sky positions (and proper motion and parallax), and $N_{\rm orb}$ is the number of independent orbital configurations.

When trying to find a timing solution for an MSP using Fermi-LAT data, the typical numbers are:
\begin{itemize}
\item Hundreds of arrival times, spread over 15 years, with significant association probabilities ($\gtrsim$\,0.2).
\item The pulsar's spin frequency and its derivatives are unknown ($N_{\rm rot} \in [10^{13}, 10^{18}]$).
\item The pulsar's position is known only to $\sim$0.1$^{\circ}$. The precision required for a phase-connected timing solution is $\sim$10$^{\text{--3}}$--1 arcseconds, depending on the rotation frequency and duty cycle. In some cases, proper motion and parallax may also be required, yielding $N_{\rm geom} \in [10^{6}, 10^{12}]$ options.
\item The binary orbit is unknown (i.e., there are 5 missing Keplerian parameters), resulting in $N_{\rm orb} \in [10^{10},10^{20}]$ different options.
\end{itemize}
Similar numbers are encountered when searching for periodicities from repeating FRBs.
The ``brute force'' way for searching a timing model is to take a group of bursts, try all combinations of parameters, compute for all arrival times their corresponding rotational phase, and perform a statistical test to detect deviations from a uniform distribution.
For short observation durations (minutes--hours), such an enumeration is feasible, and indeed, for RRATs (a special class of neutron stars, emitting pulses irregularly), this can successfully produce short-duration timing models that are phase-connected between observations using heuristic software and manual procedures~\citep{RRAT_timing}. However, this approach does not scale for long observing durations (several years) and/or when including a binary orbit (with a binary period shorter than the observation duration).
The number of options required for a complete enumeration easily exceeds $10^{30}$ for recovering a timing solution involving a binary orbit. 
Since this is unfeasible for the foreseeable future, any viable path includes an algorithmic method that is drastically different from brute-force enumeration.

The current state-of-the-art algorithms used for solving the pulsar search problem are the semi-coherent enumeration algorithms \citep{NiederSemiCoherentCircular, PletchSemiCoherent}. These algorithms utilize a special detection statistic that reduces the number of trials by reducing the coherence time, trading off the overall search sensitivity for a much reduced computational complexity.
Currently, these algorithms use one of the largest computing networks on the planet, \textit{Einstein@Home}~\citep{Allen+2013}, which spends up to 1000 core years per target. Even with the best computing resources, using state-of-the-art methods, the pulsar search problem could be solved blindly only when restricted to isolated pulsars and with reduced sensitivity (due to a relatively short coherence time used to reduce the computational load).

\subsection{Our Contributions}
In a series of papers, we cast the pulsar search problem into the problem of finding a short vector in a lattice and show the remarkable utility of lattice reduction and sieving for solving the pulsar search problem. 
A proof of concept for an algebraic algorithm that solves the pulsar search problem (in contrast to the enumeration techniques currently employed) is presented in this paper. The algorithm exactly converts the astrophysical question into the problem of finding the shortest (non-trivial) vector in a lattice. This class of algorithms has been extensively developed with cryptanalysis applications in mind. As far as we know, this is the first application of these algorithms, in high dimension and of cryptographic hardness, outside of cryptography and number theory, although it was used as a speed-up tool in satellite navigation systems \citep{teunissen1993least}.

In this paper, we first show how the pulsar detection problem on sparse data could be written as finding a short vector in a lattice. This problem, although NP-hard~\citep{SVP_NP_HARD}, is surprisingly solvable for lattices with very high dimension due to a large body of algorithms, such as the lattice reduction algorithms, LLL~\citep{LLL}, BKZ~\citep{BKZ, BKZ2}, and Gaussian Sieve algorithms~\citep{GaussSieve, bgj1_sieve}, with the most advanced algorithms combining both concepts~\citep{Sieve_dim4free, g6k}.

We then find the shortest vector in the resulting lattices using state-of-the-art off-the-shelf shortest-vector-problem (SVP) solvers \citep{fpylll, g6k}.
We demonstrate the algorithm's applicability to realistic situations using simulations and gamma-ray photon arrival times from Fermi-LAT.
Empirically and heuristically, we describe the conditions for the algorithm's success.

\subsection{Review of Existing Solutions to the Shortest Vector Problem}
A lattice $\mathcal{L}$ is the set of all linear combinations with integer coefficients of a basis of vectors (row vectors of matrix $L$, the vectors may contain any real values). 
\begin{equation}\label{eq:lattice_definition}
    \vec{v} \in \mathcal{L} \iff \exists~\vec{x}\in \mathbf{Z}^n \:\:{\rm s.t.}\:\:\:\vec{v} = \vec{x} L \new{\, ,}
\end{equation}
where $n$ is the number of basis vectors and is called the lattice dimension.
The SVP refers to the problem:
\begin{equation}\label{eq:svp_definition}
    \min_{\vec{0}\neq \vec{v}\in \mathcal{L}} {||\vec{v}||^2}.
\end{equation}
Lattice Sieving methods generate a set $V$ of many vectors of typical length $l$, and then pairs of vectors ($\vec{v}_1, \vec{v}_2 \in V$) are iteratively used to compute $\vec{w} = \vec{v}_1 - \vec{v}_2$.  If $\vec{w}$ is shorter than either $\vec{v}_1$ or $\vec{v}_2$, then it replaces them in $V$.
This process is done iteratively until convergence. If the set $V$ has more than $\sim2^{0.21n}$ vectors\new{\footnote{The exact limit is $\left(\frac{4}{3}\right)^{n/2}$, as proven by \cite{GaussSieve}.}}, then the typical lengths of vectors in the set $V$ shrink more and more until the shortest vector in the lattice is found.

Lattice reduction techniques are seeking a factorization of the lattice:
\begin{equation}
    L^t = QRU,
\end{equation} 
such that $Q$ is orthonormal, $U$ is unimodular (a matrix with determinant 1 and integer coefficients), and $R$ is upper triangular.
The reduction process gradually updates $U$ by finding ways to make the matrix $R$ more favorable for enumeration algorithms that seek to solve $R\vec{x} = \vec{v}_{\rm target}$, such as Babai's nearest plane algorithm \citep{babai1986lovasz} whose error is proportional to $\sum_i \left|R_{ii}\right|^2$.
This is accomplished by maximizing the bottom values on the diagonal of $R$.
Reduction algorithms usually achieve a matrix, $R$, with:
\begin{equation}
    \frac{R(i,i)}{R(j,j)} = \delta_{b}^{j-i},
\end{equation}
where $b$ is the block size of BKZ and $\delta_b \approx (\frac{b}{2\pi e})^{1/b}$.
The complexity of BKZ is super-exponential in $b$. 
The larger the $b$, the more convenient the enumeration. This can be intuitively understood as the greater $b$ is, the closer the values in $R$'s diagonal, which leads to easier enumeration, but this reduction is harder to achieve.
The greatest advantage of lattice reduction techniques is their ability to find very short vectors\footnote{Very short as compared to the expected shortest vector in a random lattice, with each reduction algorithm supplying some guarantees that are typically surpassed in practice.}, if such exist, even if the dimension is large.

The current state-of-the-art algorithm for solving the SVP \citep{g6k} combines both approaches, maintaining a database of short vectors on gradually increasing sub-lattices (using Babai's nearest plane algorithm \citep{babai1986lovasz}, utilizing a reduced basis), repeatedly keeping them short (using a sieve), and exploiting the database's size for an eventual sieve to solve a few extra dimensions for free \citep{Sieve_dim4free}.

\section{Pulsar Detection as Short Vector in a Lattice - The Basic Construction}
\label{sec:pulsar_as_svp_basic}

We first introduce the basic timing model used in our framework:
\begin{align}
\label{eq:basic_timing_model_p}
t_i = (K_i + \phi + \epsilon_i)P \, ,
\end{align}
where $t_i$ is the $i^{\text{th}}$ TOA, $P$ is the pulsar's rotational period, $K_i$ is an integer number of pulsar rotations since the $t=0$ reference time, and $\epsilon_i$ is the pulse phase at which this photon arrived (in the interval [-0.5, 0.5]). Assuming a Gaussian pulse profile with standard deviation $\sigma$ (where $\sigma$ is proportional to the ``duty cycle''), $\epsilon_i \sim N(0,\sigma)$.
By rearranging Equation~(\ref{eq:basic_timing_model_p}), we obtain:
\begin{equation}
    \label{eq:basic_timing_model_f}
    ft_i - K_i - \phi = \epsilon_i \, ,
\end{equation}
where $f=1/P$ is the rotational frequency of the pulsar.
Since we assume the pulse profile is of Gaussian shape, according to the Neyman-Pearson lemma~\citep{neyman_pearson}, the strongest statistical test to decide between the null hypothesis, uniform phase residuals, and the alternative Gaussian pulse profile is
\begin{equation}\label{eq:residuals_sum}
    \mathcal{T} = \sum_i \epsilon_i^2 \, .
\end{equation}
For specific values of $f$, $\phi$, and ${K_i}$ to indicate a potential detection, they should first be better than the alternative values. So even before considering the significance, we need to find the best values for the parameters, which requires minimizing $\mathcal{T}$.
To prevent undesired solutions of extremely high frequencies (for example, due to the clock frequency of the detector), we also enforce a Gaussian prior on the parameters:
\begin{align}\label{eq:svp_basic_priors}
    V\left[\frac{f}{f_{\rm prior}}\right] = 1, \nonumber \\
    V\left[\phi\right] = 1,
\end{align}
when adding them to the test statistic, we will ensure each of those priors contributes $\sigma^2$, the same as any other coordinate
\begin{equation}\label{eq:test_statistic}
    \mathcal{T} = \sum_i \epsilon_i^2 
    + \left(\sigma\frac{f}{f_{\rm prior}}\right)^2 
    + \left(\sigma\phi\right)^2\, .
\end{equation}
The lattice structure arises from the restriction that the $K_i$'s are integers, $f$ and $\phi$ are unknown and of some precision.
The equation needs to be strictly linear in the unknowns and all unknown coefficients must be strictly integers, as required by the lattice solver that we use (G6K;~\citealt{g6k}). For this purpose, Equation~(\ref{eq:basic_timing_model_f}) can be re-expressed as:
\begin{align}\label{Eq:BasicLatticeEq}
    \nint*{\frac{f}{d_f}} d_f t_i - K_i - \nint*{\frac{\phi}{d_\phi}} d_\phi = \epsilon_i
\end{align}
Here, $d_f$ is the measurement resolution of $f$ in the integer solution, which we choose following
\begin{align}
    d_f \ll \frac{\sigma}{(\max_i t_i - \min_i t_i)} \,,
\end{align}
to ensure the rounding resolution is much smaller than the expected parameter's precision to avoid rounding errors.
Writing the lattice basis vectors as {\bf rows} of the following matrix, we can write\footnote{The G6K lattice solver~\citep{g6k} also requires that the input matrix's entries are integers, so we multiply the elements by a large number and round. This number is chosen to be much larger than the largest expected solution coefficient divided by the phase residual's resolution, typically $\gg$\,$\frac{f_{\rm prior}}{d_f \sigma^2}$.}:
\begin{align}
L_{\rm per} = 
\begin{pmatrix}
1   &  0  & 0 &  \dots  &  0  & 0 & 0 \\
0   &  1  & 0 &  \dots  &  0  & 0 & 0 \\
0   &  0  & 1 &  \dots  &  0  & 0 & 0\\
\vdots & \vdots & & \dots & \vdots & \vdots & \vdots\\
0   &  0  & 0 &  \dots  &  1  & 0 & 0 \\
d_ft_0 & d_ft_1 & d_ft_2 &\dots   & d_ft_n & \eta_t & 0 \\
d_\phi & d_\phi & d_\phi &\dots   & d_\phi & 0      & \eta_\phi \\
\end{pmatrix} \, ,
\end{align}
where we introduced $\eta_t, \eta_\phi$ to account for the priors in Equation~(\ref{eq:test_statistic}), and a choice for them that will give the same expected loss in every coordinate is:
\begin{align}
\eta_t = \sigma \frac{d_f}{f_{\rm prior}} \,,\:\:\:\: \eta_\phi = \sigma d_\phi.
\end{align}
Using this lattice basis definition, the vectors in the lattice will have the components of our test statistic from Equation~(\ref{eq:test_statistic}) as their entries, and its length that we minimize when searching for the shortest vector corresponds to the test statistic, incorporating the phase residuals and priors.

\section{Extending the Lattice-based Solution to other unknowns}
\label{sec:pulsar_as_svp_more}

We have described the technique for using the lattice framework to find simple, perfectly periodic solutions. However, the full advantage of using the lattice approach is incorporating many other timing parameters into the model without significantly increasing the computational complexity.
The most prominent of these are the pulsar spin-down parameters ($\dot{f}$, $\ddot{f}$, ...), the barycentric-correction parameters, and some of the Keplerian orbit parameters.

\subsection{Spin-down Parameters}
The spin frequency of rotating neutron stars can change due to processes such as magnetic breaking and accretion.
We can calculate the rotational phase of each TOA, $t_i$, using:
\begin{align}
    \phi + K_i + \varepsilon_i = \int_{t_{\rm ref}}^{t_i}{f(t)dt} \;,
\end{align}
where $K_i$ are integers, $t_{\rm ref}$ is the reference time, and $f(t)$ is the instantaneous rotation frequency. 
Expanding $f(t)$ as a Taylor series and integrating, the arrival times satisfy the equation:
\begin{align}
 K_j + \epsilon_j = \phi + ft_j + \sum_{k=2}^m{\frac{f^{(k-1)}}{k!} t_j^k}.
\end{align}
Similarly to Equation~(\ref{Eq:BasicLatticeEq}), we can equivalently write:
\begin{align}
K_j + \epsilon_j = \phi + \left[\frac{f}{d_f}\right]d_ft_j + \sum_{k=2}^m{\left[\frac{f^{(k-1)}}{d_{f^{(k)}}k!} \right] d_{f^{(k-1)}}t^k}.
\end{align}
This form of the equation can be naturally encapsulated by adding a few additional vectors to the lattice:
\begin{align}
L_{\rm spin-down} = 
\begin{pmatrix}
I_{n\times n} & \mathbb{0}_{n\times (m+1)}\\
S_{(m+1)\times n} & {\eta}_{(m+1)\times (m+1)}\\
\end{pmatrix},
\end{align}
where $I_{n\times n}$ is the identity matrix, $\mathbb{0}_{n\times (m+1)}$ is a zero matrix, $S_{(m+1)\times n}$ is a matrix containing the spin-down vectors, and ${\eta}_{(m+1)\times (m+1)}$ is a diagonal matrix. $S_{(m+1)\times n}$ has the form:
\begin{align}
S_{(m+1)\times n} = 
\begin{pmatrix}
d_\phi                            & \dots   & d_\phi                                \\
d_f t_0                           & \dots   & d_f t_{n-1}                           \\
\frac{1}{2} d_{f^{(1)}}(t_0)^2    & \dots   & \frac{1}{2} d_{f^{(1)}}(t_{n-1})^2    \\
\vdots                            & \ddots  & \vdots                                \\
\frac{1}{m!} d_{f^{(m-1)}}(t_0)^m & \dots   & \frac{1}{m!} d_{f^{(m-1)}}(t_{n-1})^m \\
\end{pmatrix} \,,
\end{align}
and the elements of ${\eta}_{(m+1)\times (m+1)}$ are given by:
\begin{align}
\eta(1,1) &=  \frac{\sigma}{d_\phi} \,,\\
\eta(k+1,k+1) &= \sigma \frac{d_{f^{(k)}}}{f^{(k)}_{\max}} \,.
\end{align}

\subsection{Accounting for an Unknown Position}

The Fermi satellite follows Earth's orbital motion, which, if uncorrected, will leave a 500-second differential residual in the photon arrival times.
The main component of the time correction is to project the satellite's position vector relative to the solar system's barycenter along the direction of the source.\footnote{There are also general relativistic propagation effects and motions due to other planetary bodies in the solar system that need to be corrected. This can be accounted for, for example, using the barycentering routines available in the PINT software package~\citep{PINT2}.} Since the required time resolution for efficient MSP recovery is of order 10$^{\text{--4}}$\,s, the sky position needs to be known to a precision that is better than $\frac{10^{-4}}{500} \,{\rm rad}$\,$\approxeq$\,10\,mas. The localization precision of sources in the 4FGL~catalog is roughly 0.1$^{\circ}$.
Thus, for pulsar searches, there are 10$^{\text{9}}$ different trial positions that are possible in a blind search.

Moreover, the proper motion of the pulsar can also substantially affect the timing solution (especially for MSPs) because the resolution required by the proper motion is of order $1 {\rm \frac{{\rm mas}}{year}}$.
A pulsar that is located at a distance of 1\,kpc, with a tangential velocity of 100\,km/s, will have a proper motion of order 10\,mas/year, which will introduce substantial timing residuals.
To solve for the pulsar's precise position with the lattice, we notice that the space of all possible corrections is linear (although the coordinate transformation between this space and the commonly used coordinates is not completely linear).
We collectively denote all of the positional parameters as $\vec{\psi}$, and their collective phase delay (time delay times frequency) by $F(\vec{\psi})$.
We can then write:
\begin{align}
&f t_j = K_j + F(t_j,\vec{\psi}) = \nonumber\\
&K_j + F(t_j,\vec{\psi_0}) + (\vec{\psi}-\vec{\psi}_0) \frac{dF}{d\vec{\psi}}(t_j,\vec{\psi_0})
\end{align}
This linearized model for the timing solution could be added to the lattice via the same procedure described for spin-down parameters.
In general, this approximation corresponds to trying to reconstruct a 3-parameter function with non-linear constraint (a unit vector pointing at the pulsar's direction) using 2 unconstrained parameters. We might naively expect the approximation error to grow as
\begin{equation}
    \Big|(\vec{\psi}-\vec{\psi}_0)_l \frac{d^2F}{d\psi_ld\psi_m}|_{(t_j,\vec{\psi_0})}(\vec{\psi}^\prime-\vec{\psi}_0)_m\Big| \, ,
\end{equation}
and result in an intolerable error (barycentric corrections that are nonphysical) for relatively small position offsets. However, the Fermi satellite's motion around the solar system's barycenter is nearly planar, and if we look in ecliptic coordinates, we need only 2 weakly constrained parameters to linearly combine $(x_{\rm Fermi}(t_i), y_{\rm Fermi}(t_i))$. This means that even large offsets will still yield a barycentric correction corresponding to some pulsar position.
Also, the mixing between the barycentric correction and the first frequency derivative is negligible for reasonable search parameters\footnote{$\varepsilon=\frac{1}{2}\dot{f}\left(2T\Delta\alpha\frac{AU}{c}\right)\approx 3\cdot10^{-3}\left(\frac{\dot{f}}{-10^{-11}\text{s}^{-2}}\right)\left(\frac{\Delta\alpha}{0.1\arcdeg}\right)$}.
Therefore, there is no need for external enumeration of the exact position of the pulsar, and our linear approximation is sufficient.

\subsection{Adding a Circular Orbit into the Lattice}

Many of the pulsars that we aim to detect (for example, most MSPs) are in binary systems. Therefore, we construct a feasible algorithm for blind detection for such systems.

This issue was not addressed in the previous blind pulsar search surveys because brute-force enumeration of all of the orbital parameters, along with the spin frequency, spin frequency derivatives, sky position, and proper motion, is unfeasible. Brute-force enumeration of all of these parameters can easily accumulate to more than $10^{30}$ independent options in the parameter space.

In Section \ref{sec:fermi_demo}, we demonstrate that with the lattice-based solution, we can solve for position and spin-down in $\sim 100$ core-seconds, which, even with the simple, brute force approach, calls to reconsider the binary search problem.

The lattice-based solution efficiently solves linear integer least squares problems. 
Therefore, to solve for the orbit, we must also linearize the phase space of all circular orbits as much as possible.
The orbital reference phase and semi-major axis are trivially linearizable:
\begin{align}
    &v_{1,\rm orb} = \sin(\Omega_{\rm orb} t) \nonumber\\
    &v_{2,\rm orb} = \cos(\Omega_{\rm orb} t)
\end{align}

The orbital frequency, $\Omega_{\rm orb}$, is more challenging, as periods that differ by an integer number of orbits during the observation duration are approximately orthogonal. Therefore, we must divide the parameter space into a union of many different linear spaces, covering all of the options for $\Omega_{\rm orb}$. For a circular orbit with an orbital period of $10$ hours, a semi-major axis of $1$ light second, an observation duration of 10 years, and a target timing precision of $0.1\,$ms, this amounts to $\frac{a}{\sigma P}\frac{T}{P_{\rm orb}}\approx 10^{8}$ different period trials. Since each trial requires solving the SVP problem (at least 1--100 CPU seconds, depending on dimension, see~\citealt{ducas2021advanced}), this will be extremely demanding with an estimated cost of hundreds of core years. 
We can reduce this number of trials by adding the following vectors to the lattice:
\begin{align}
    &v_{3,\rm orb} = t\sin(\Omega_{\rm orb} t) \nonumber\\
    &v_{4,\rm orb} = t\cos(\Omega_{\rm orb} t)
\end{align}
These vectors were obtained by taking the derivative of the orbital time delay with respect to $\Omega_{\rm orb}$. Incorporating these vectors significantly reduces the number of required orbital period trials. For example, for the aforementioned parameters, the number of orbital period trials decreases from approximately $10^{8}$ to approximately $\sqrt{\frac{a}{\sigma P}}\frac{T}{P_{\rm orb}}\approx10^{6}$.
We can add more derivatives to the lattice, reducing the required enumeration. 
Unlike adding the first derivatives, when we add the second derivatives, the sensitivity can be compromised because not every point in the lattice is physical, and this can artificially increase the look-elsewhere effect.
This is because the coefficients of the second derivative vectors are fully determined, in a non-linear way, from the coefficients of the first. For example, the ratio between the coefficients of $v_{1,\rm orb}$ and $v_{2,\rm orb}$ is the same as between the coefficients of $v_{3,\rm orb}$ and $v_{4,\rm orb}$, because $v_{4,\rm orb}$ is a second derivative vector, with respect to the phase and orbital frequency.

Note that the spin-down vectors (corresponding to $(f^{(1)},f^{(2)},\dots)$) are correct when using the source time, which is inaccessible to us (we know only the observed time). This introduces a coupling between the spin-down parameters and the orbital parameters. Fortunately, the two vectors correcting these coefficients are the same as those compensating for orbital period change and small eccentricity changes.

An in-depth discussion on partitioning the enumeration space (including a full Keplerian orbit) into a set of linear spaces will be described in future work. In this paper, we discuss the case of a pulsar in a circular orbit.

\section{Information Limit - Performance under the Null Hypothesis $H_0$}
\label{sec:h0}

Understanding the expected performance without a signal is important when using this lattice-based solution as a detection tool. Therefore, we compute the expected length of the shortest non-trivial vector in a lattice constructed with random TOAs.
A useful tool to analyze our lattice setup is the Gaussian heuristic (GH), which states that the probability of finding lattice sites in some volume is proportional to the volume, and the expected length (per coordinate) of the shortest vector in a lattice is:
\begin{equation}
    \lambda_1=\frac{\text{vol}(\mathcal{L})^{1/n}}{\sqrt{2\pi e}}\,,
    \label{eq:GH}
\end{equation}
where $n$ is the lattice dimension and $\text{vol}(\mathcal{L})$ is the lattice volume, calculated as:
\begin{equation}
    \text{vol}(\mathcal{L}) = \sqrt{\det LL^T}\,.
    \label{eq:lattice_volume}
\end{equation}
But, because the original lattice's volume is wholly controlled by $d_\phi, d_f, \dots$, the arbitrarily small numbers intended to make the timing vectors quasi-continuous, the GH is unsuitable for analyzing it.

Note that once we set the coefficients for the unit vectors (or the number of integer revolutions for each TOA), we only need to perform a simple linear fit with respect to the timing vectors to find the smallest phase residuals. But, we can reverse the order, orthogonalize the unit vectors with respect to the timing vectors, and then perform the integer search.
We will refer to this reverse order search as searching in the sub-lattice orthogonal to the timing vectors, and it is suitable for analysis using the GH\footnote{Note, in solutions spanned by the orthogonal sub-lattice, some of the phase residuals might be bigger than half a revolution, $|\varepsilon_i|>0.5$.}.
We use the following lattice:
\begin{equation}
    L = 
    \begin{pmatrix}
    I_{n\times n} & \mathbb{0}_{n\times m}\\
    V_{m\times n} & {\eta}_{m\times m}\\
    \end{pmatrix} \, ,
\end{equation}
where
\begin{equation*}
    V_{m\times n}^T = 
    \begin{pmatrix}
        d_1 \vec{v}_1 & \cdots & d_m \vec{v}_m
    \end{pmatrix} \, ,
\end{equation*}
and
\begin{equation*}
    \left(\eta_{m\times m}\right)_{i,j} = \delta_{i,j} d_i \sigma_{\rm exp} / \sigma_i \, ,
\end{equation*}
where $\vec{v}_i$ are the orthonormalized timing vectors and $d_i$ are small factors used to make the timing model vectors $\vec{v}_i$ arbitrarily small. $\sigma_{\text{exp}}$ is the expected length of a random vector (which we will now solve for self-consistently), and $\sigma_i$ is the range that we search for in $v_i$ (similarly to $f_{\rm max}$ for the periodicity vector).
Following these definitions, we can calculate the lattice volume (see Appendix \ref{app:volume_calculation} for a detailed calculation):
\begin{equation}
\label{eq:h0_volume}
    \text{vol}(\mathcal{L}) = \prod_i \left( \frac{\sigma_{\text{exp}}/\sigma_i}{\sqrt{1 + \left(\sigma_{\text{exp}}/\sigma_i\right)^2}} \right) \, .
\end{equation}
Next, after substituting Equation~(\ref{eq:h0_volume}) into Equation~(\ref{eq:GH}), we obtain:
\begin{align}
    \sigma_{\text{exp}} \equiv \lambda_1 
    &= \left(\prod_i \frac{\sigma_{\text{exp}}/\sigma_i}{\sqrt{1 + \left(\sigma_{\text{exp}}/\sigma_i\right)^2}}\right)^{1/n} / \sqrt{2\pi e} \, .
\end{align}

Solving for $\sigma_{\text{exp}}$ generally involves solving a high-degree polynomial and requires a numerical treatment. But, in the typical case, we search for solutions with large ranges for the timing vectors (relative to a single phase cycle), allowing us to simplify the calculation:
\begin{equation}
    \label{eq:sigma_exp_approx}
    \sigma_{\text{exp}}
    \approx \sigma_{\text{exp}}^{m/n} \left(\prod_i\sigma_i\right)^{-1/n} / \sqrt{2\pi e} \, ,
\end{equation}
After solving Equation~(\ref{eq:sigma_exp_approx}) for $\sigma_{\text{exp}}$, we obtain:
\begin{equation}
    \sigma_{\text{exp}}\approx
    \left(\prod_i\sigma_i\right)^{-\frac{1}{n-m}}/\sqrt{2\pi e}^{\frac{n}{n-m}} \, .
\end{equation}

Knowing how to compute the minimum length of a spurious signal precisely, we can estimate the false alarm probability (FAP) of a candidate signal with $\sigma_{\rm cand}$ by computing the expected number of lattice sites with $\sigma \le \sigma_{\text{cand}}$ using the~GH. The estimated FAP is given by: 
\begin{align}
    {\rm FAP} = \left(\sigma_{\rm cand} / \sigma_{\rm exp}\right)^{n} \,.
\end{align}

\section{Complexity Analysis}
\label{sec:complexity}

Following the same logic, we can compute the complexity (and the amount of data) needed to solve a timing problem with a given entropy (number of ``independent'' options):
\begin{align}
    \Lambda &\equiv \frac{1}{\text{vol}(\mathcal{L})} \nonumber\\
    &=\prod_i \frac{\sigma_i}{\sigma_{\rm exp}} \sqrt{1 + \left(\sigma_{\rm exp}/\sigma_i\right)^2} \, .
\end{align}
In the case where the correct solution isn't the shortest vector
\begin{equation}
    \sigma_{\rm exp} < \sigma \,,
\end{equation}
we might still be able to find it by generating $N_{\rm candidates}$ short vectors, while keeping the condition
\begin{equation}
    \left(\frac{\sigma}{\sigma_{\rm exp}}\right)^n \le N_{\rm candidates} \,.
\end{equation}
For the lattice sieve, we use $N_{\rm candidates}\approx 2^{0.2n}$ and the time complexity\footnote{Time complexity refers to the number of basic computer operations required.} is $C(n)\approx2^{0.36n}$, as measured by \cite{ducas2021advanced} for $n\sim100$.

\begin{deluxetable}{ccccc}
\tabletypesize{\footnotesize}
\tablewidth{0pt}
\tablecaption{Complexity of the lattice sieve for different pulse widths and corresponding association probabilities. \label{tab:complexity}}
\tablehead{
\colhead{$\sigma$} & \colhead{$p(\rm at \; \sigma_{\rm int}=0)$}& \colhead{$C(\Lambda)$} & \colhead{$C(\Lambda=10^{28})$} &
\colhead{$n(\Lambda=10^{28})$}
}
\startdata 
$0.2$  & $0.5$  & $\Lambda^{0.81}$ & $10^{22.5}$ & $207$\\
$0.16$ & $0.7$  & $\Lambda^{0.44}$ & $10^{12}$   & $111$\\
$0.11$ & $0.85$ & $\Lambda^{0.27}$ & $10^{7.5}$  & $69$\\
$0.06$ & $0.95$ & $\Lambda^{0.17}$ & $10^{4.7}$  & $42$\\
\enddata
\tablecomments{$\sigma$ is the pulse width, $p$ is the association probability of the TOAs, $\sigma_{\rm int}$ is the intrinsic pulse width, and they are related by $\sigma^2 = p\cdot \sigma_{\rm int}^2 + (1-p)\cdot \frac{1}{12}$.}
\end{deluxetable}

In Table \ref{tab:complexity}, we provide the time complexity of our method for different pulse widths and their corresponding association probabilities. We find that $\sigma<0.11$ is easily attainable, while $p=0.5$ is still unfeasible in this framework, even with an intrinsically infinitely narrow pulse.

\section{Injection Recovery - Performance under the Alternative Hypothesis $H_1$}
\label{sec:h1}

As in any detection problem, we need to analyze our algorithm in the presence of a signal, and in this case, we are presented with a challenge. 
Since we use the G6K lattice solver, which is designed to solve the~SVP and not the shortest non-trivial vector problem by enumeration, the solver may not always find the shortest non-trivial vector.
Therefore, we performed a basic injection-recovery analysis of an isolated MSP with a characteristic age of 100\,Myr and a position known to a precision of 1$^{\circ}$ ($\Lambda_d\sim10^{28}$).

\begin{figure*}
    \gridline{\fig{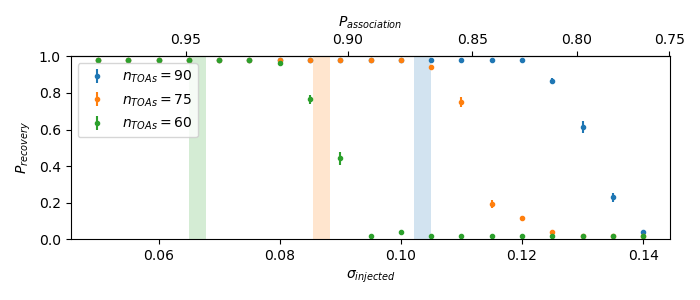}{0.8\textwidth}{}}
    \gridline{\fig{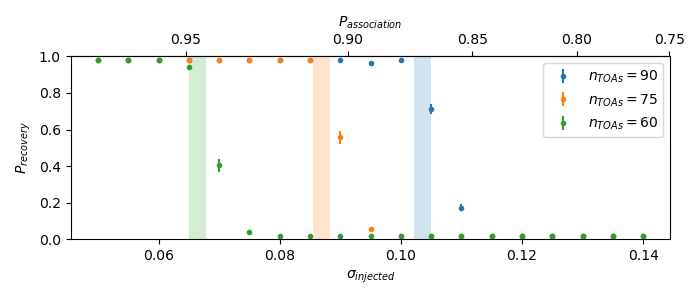}{0.8\textwidth}{}}
    
    \caption{The probability for recovery as a function of the injected signal's width $\sigma$. Top: Fermi-LAT-like simulation with an additional verification step. Bottom: FRB-like simulation with no additional verification step.
    Recovery probability was calculated based on 50 injections per $\sigma$ for different numbers of TOAs over 10 years. The simulation consisted of an isolated MSP ($f \in [100,1000]$\,Hz and $\tau$ = 100\,Myr). The expected information limit, $\sigma_{\text{exp}}$ from Section \ref{sec:h0}, is indicated using the shaded vertical stripes. The information limits appear as a stripe, rather than a single line, because the entropy depends on the specific simulated sky position.
    A sharp transition from constant to no recovery at and after $\sigma_{\text{exp}}$ is seen in both the FRB-like and Fermi-LAT-like simulations.
    An important distinction between the two simulations is that we check whether the solution is in the short vectors set in the Fermi-LAT-like simulation, while we check whether the solution is the \textit{shortest} in the short vectors set in the FRB-like simulation.
    \label{fig:inj-rec}}
\end{figure*}

We performed this analysis in both an FRB-like and a Fermi-LAT-like scenarios.
In the FRB-like scenario, we sieve in the sub-lattice that contains all non-continuous vectors (not $d_f\vec{t}$, $d\vec{\phi}$, etc...) and compare its shortest non-trivial vector against the injected one. The results, shown in Figure~\ref{fig:inj-rec}, demonstrate that the $H_0$ limit can be reached but not surpassed.

In the Fermi-LAT-like scenario, we use only photons with high association probabilities in the lattice and later verify the solutions using photons with lower association probabilities. Therefore, we sieve to generate many candidate solutions (in the same sub-lattice as in the FRB scenario). If the injected signal is one of the candidates, we regard it as a successful recovery. The results from this analysis (see Figure~\ref{fig:inj-rec}) show that it is possible to significantly surpass the $H_0$ limit because we allow multiple trials.

\section{Results on Real Data} 
\label{sec:fermi_demo}

\begin{figure}
    \centering
    \includegraphics[width=0.45\textwidth]{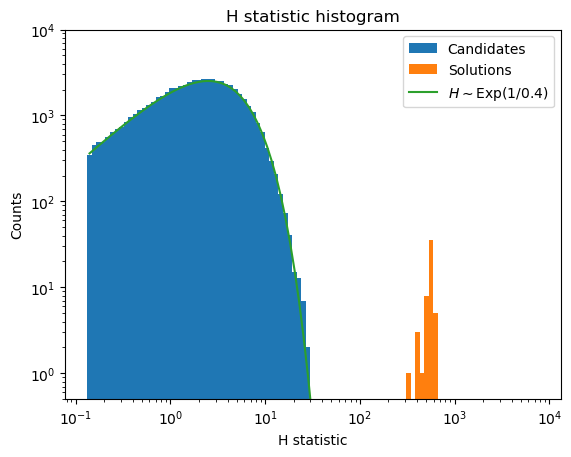}
    \caption{The H-test statistic histogram for significant and insignificant candidates (which we will call solutions and candidates, respectively) generated using the lattice sieving. The H-test is calculated over the verify-set of photons, and the significance threshold for the p-value is set at $10^{-7}$. The green line corresponds to the distribution of an exponential random variable with parameter $\lambda=1/0.4$, as expected from the simulation by ~\citealt{Htest_distribution_function}.}
    \label{fig:h_stat}
\end{figure}

\begin{figure}
    \centering
    \includegraphics[width=0.45\textwidth]{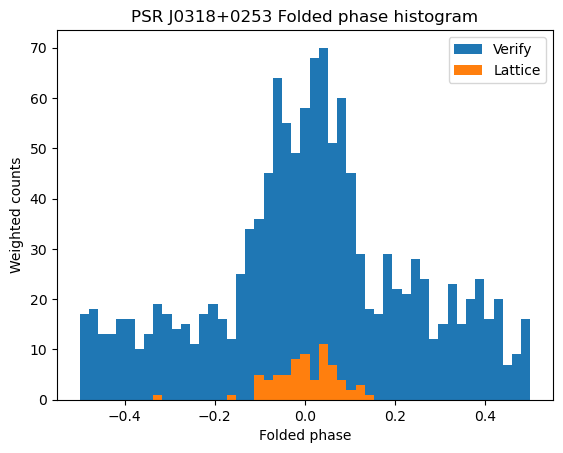}
    \caption{Weighted histograms of one of the solution's phase fold for the lattice-set and verify-set photons.}
    \label{fig:fold}
\end{figure}

Here, we present the application of this method to \textbf{4FGL~J0318.2+0254}, a Fermi-LAT source also known as \textbf{PSR J0318+0253}, discovered using the FAST radio telescope~\citep{3FGLJ0318.1+0252}. 4FGL~J0318.2+0254 is an isolated MSP with a sufficient number of high probability photons and a narrow enough pulse.
We used the Fermi-LAT data from 31 July 2008 to 28 September 2023 and selected photons arriving within 3$^{\circ}$ of the source's position in 4FGL-DR4~\citep{4FGL-DR4}. We assigned association probabilities to the photons using the standard fermitools procedure~\citep{FERMITOOLS}.
We divided the photons into two sets based on their association probabilities: the 70 highest probability photons and the rest with a lower limit of $p\ge 0.2$, which we now refer to as the lattice-set and verify-set, respectively.
We construct the lattice using the lattice set of high probability photons, searching for $f$\,$\sim$\,100\,Hz and $\tau$\,$\sim$\,100\,Myr, assuming the position reported in 4FGL-DR4 (with errors equal to the $95\%$ confidence interval uncertainties) and a proper motion of the order 10\,mas/yr.
We reduced the lattice and generated $\sim6.5\times10^4$ short candidates. From these candidates, we selected the ones with high enough $f$ and used them to fold the verify-set, calculating the H-test statistic~\citep{HTEST} for each candidate. The results from this analysis are shown in Figure~\ref{fig:h_stat}. Using the H-test scores of the verify-set of photons, we identified the correct solutions. The phase folded histogram from one of these correct solutions is shown in Figure~\ref{fig:fold}.\footnote{A notebook that roughly follows this procedure is available at: \url{https://github.com/Zackay-Lab/pulsar-lattice-example}.}

\section{The Norm Problem - Adapting to Bad Photons and Double Pulse Profiles}
\label{sec:conceptual_problems}

The most severe conceptual problem in our setup is the fact that we are using the $L_2$ norm to decide between the different possible solutions. 

The $L_2$ norm corresponds to the assumption of a perfect Gaussian pulse profile, which doesn't hold for a vast majority of known $\gamma$-ray pulsars that tend to have a double pulse profile. This is also a bad norm to use when background photons are present, the typical case for sources in Fermi-LAT. 

A better choice of a test statistic to rank the different lattice vectors is the H-test~\citep{HTEST}. The H-test corrects for both a somewhat more general pulse shape and for the fact that different TOAs have different association probabilities with the source (at least in the Fermi-LAT case). 

Moreover, it is useful to output many vectors from the lattice sieve and rank them according to the more sensitive H-test, thereby picking up the correct solution even if the $L_2$ norm ranks the correct solution only in the $M^{\text{th}}$ place. Sieving algorithms are usually using a large number of vectors, and this approach is beneficial to increase the sensitivity of the search. 

Additionally, the requirement for a large number ($N$\,$>$\,60) of photons with high association probability ($p>0.85$) limits the applicability of the method to a few dozen sources (out of the thousands of unassociated sources).
In a follow-up paper, we will present an algorithmic improvement that will more effectively recover the correct solution in the situation of only moderate association probabilities ($p$~$\sim$~0.5).

\section{Conclusions}
\label{sec:conclusions}

In this paper, we have shown that the timing solutions of pulsars can be recovered using lattice algorithms, a set of advanced tools developed for cryptanalysis. We have also demonstrated that a lattice-based approach can be used to solve a timing model that is a linear combination of known vectors.
Additionally, we showed that lattice algorithms can solve problems that were previously impossible to solve, in a matter of seconds.

We discussed the computational complexity of the lattice algorithm technique and showed that it is substantially smaller than full enumeration of the parameter space. The results strongly depend on the duty cycle (or the equivalent width, for more complex pulse shapes) and the association probabilities of the arrival times.

As a proof-of-concept, we recovered the timing solution of a real pulsar using Fermi-LAT data and our lattice algorithm implementation.
In the next papers in this series, we will show how to linearize a Keplerian orbit and present a novel algorithm that is more computationally efficient, allowing us to solve lattice problems in which half of the TOAs are random or cases where the pulsar has a double pulse profile. We will then apply our methods to all relevant Fermi-LAT unassociated sources.

\section*{Acknowledgements}
\label{sec:ack}

A.B.P. is a Banting Fellow, a McGill Space Institute~(MSI) Fellow, and a Fonds de Recherche du Quebec -- Nature et Technologies~(FRQNT) postdoctoral fellow. B.Z. is supported by a research grant from the Willner Family Leadership Institute for the Weizmann Institute of Science. 

This research was supported by the Minerva Foundation with funding from the Federal German Ministry for Education and Research. This research was also partially supported by the Israeli Council for Higher Education~(CHE) via the Weizmann Data Science Research Center.

We greatly thank Tejaswi Venumadhav, Liang Dai, Matias Zaldarriaga, Martin Albrecht, Leo Ducas, Adi Shamir, Zvika Brakerski, Ittai Rubinstein, Asaf Shuv, Nathan Keller, Noga Bashan, and Doron Walach for in-depth discussions over various aspects of this project.

\appendix

\section{Lattice volume calculation}
\label{app:volume_calculation}

According to the Gaussian heuristic (GH), the expected length of the shortest vector in a lattice is (per coordinate):
\begin{equation}
    \lambda_1=\frac{\text{vol}(\mathcal{L})^{1/n}}{\sqrt{2\pi e}}\,,
    \label{eq:GH_app}
\end{equation}
where $n$ is the lattice dimension and $\text{vol}(\mathcal{L})$ is the lattice volume, calculated as:
\begin{equation}
    \text{vol}(\mathcal{L}) = \sqrt{\det LL^T}\,.
    \label{eq:lattice_volume_app}
\end{equation}
In our setup, we are interested in the sub-lattice consisting of the unit vectors of the different TOAs, projected orthogonally to the quasi-continuous timing model vectors, the phase residuals lattice.
Using the following lattice:
\begin{equation}
    L = 
    \begin{pmatrix}
    I_{n\times n} & \mathbb{0}_{n\times m}\\
    V_{m\times n} & {\eta}_{m\times m}\\
    \end{pmatrix} \, ,
\end{equation}
where
\begin{equation*}
    V_{m\times n}^T = 
    \begin{pmatrix}
        d_1 \vec{v}_1 & \cdots & d_m \vec{v}_m
    \end{pmatrix} \; ; \; \vec{v}_i\cdot\vec{v}_j=\delta_{ij} \, ,
\end{equation*}
and
\begin{equation*}
    \left(\eta_{m\times m}\right)_{i,j} = \delta_{i,j} d_i \sigma_{\rm exp} / \sigma_i \, ,
\end{equation*}
where $\vec{v}_i$ are the orthonormal timing vectors, $d_i$ are small constants, $\sigma_i$ are the extent of the timing vectors we wish to search over, and $\sigma_{\rm exp}$ is the length per coordinate of the shortest random vector we are trying to estimate.
We wish to project out the $\begin{pmatrix} V_i & \eta_i \end{pmatrix}$ directions, \new{and for that, it would be useful to define the unscaled versions of $V$ and $\eta$
\begin{equation}
    U^T\equiv
    \begin{pmatrix}
        \vec{v}_1 & \cdots & \vec{v}_m
    \end{pmatrix}
    \;;\;
    \Sigma\equiv\delta_{i,j}\sigma_{\text{exp}}/\sigma_i.
\end{equation}}
This can be done by using the projection operator:
\begin{align}
    P_V &= I_{n+m\times n+m} - \begin{pmatrix} U & \Sigma \end{pmatrix}^T \left( \begin{pmatrix} U & \Sigma \end{pmatrix} \begin{pmatrix} U & \Sigma \end{pmatrix}^T\right)^{-1} \begin{pmatrix} U & \Sigma \end{pmatrix} \, ,
\end{align}
the projected sub-lattice can then be written as:
\begin{equation}
    \Pi_{n\times n+m} = I_{n\times n+m} P_V \, ,
\end{equation}
and the matrix we are interested in its determinant is:
\begin{align}
    \Pi \Pi^T
    &= I_{n\times n+m}P_V P_V I_{n+m\times n}\nonumber\\
    &= I_{n\times n+m}P_V I_{n+m\times n}\nonumber\\
    &= I_{n\times n} - U^{T} \left( I_{m\times m} + \Sigma \Sigma^T \right)^{-1} U \,.
\end{align}
Now, to calculate the determinant, we use the Weinstein–Aronszajn identity:
\begin{align}
    \det \Pi \Pi^T
    &= \det I_{n\times n} - U^{T} \left( I_{m\times m} + \Sigma \Sigma^T \right)^{-1} U \\
    &= \det I_{m\times m} - \left( I_{m\times m} + \Sigma \Sigma^T \right)^{-1} U U^{T} \\
    &= \det I_{m\times m} - \left( I_{m\times m} + \Sigma \Sigma^T \right)^{-1} \, ,
\end{align}
if $\Sigma$ is diagonal, this reduces to
\begin{equation}
    \det \Pi \Pi^T = \prod_i \frac{\Sigma_i^2}{1 + \Sigma_i^2} \, ,
\end{equation}
and the volume is
\begin{equation}
    \text{vol}(\mathcal{L}) = \prod_i \frac{\Sigma_i}{\sqrt{1 + \Sigma_i^2}} \, .
\end{equation}

\bibliographystyle{aasjournal}
\bibliography{larp.bib}

 \end{document}